\documentstyle[12pt]{article}

\newcommand{\beq}{\begin{equation}}
\newcommand{\eeq}{\end{equation}}
\newcommand{\bea}{\begin{eqnarray}}
\newcommand{\eea}{\end{eqnarray}}
\newcommand{ \nn}{\nonumber}

\newcommand{\gtrsim}{\ \rlap{\raise 2pt\hbox{$>$}}{\lower 2pt \hbox{$\sim$}}\ }
\newcommand{\lessim}{\ \rlap{\raise 2pt\hbox{$<$}}{\lower 2pt \hbox{$\sim$}}\ }

\newcommand{\prl}[1]{{\it Phys. Rev. Lett.} {\bf #1}}

\begin{document}
\pagestyle{empty}
\begin{titlepage}
\today
\begin{center}
\vspace*{-2cm}
\hfill FTUV/97-58  \\
\hfill IFIC/97-88  \\
\vspace{1cm}
{\Large \bf P-odd observables at the $\Upsilon$ peak } \\
\vspace{1cm}

{\large J. Bernab\'eu}$^a${\large, F.J. Botella}$^{a,b}${\large, O. Vives}$^
{a,b}$\\
\vspace*{0.3cm}
$^a$Departament de F\'{\i}sica Te\`orica, Universitat de Val\`encia. \\
$^b$I.F.I.C. (Centre Mixte Universitat de Val\`encia--C.S.I.C.) \\
E-46100 Burjassot (Val\`encia, Spain) 
\vspace*{0.5cm} \\
\begin{abstract}
We study the $\gamma$-Z interference in the process $e^+ e^-\rightarrow
\Upsilon \rightarrow \tau^+ \tau^-$ as a means to measure the neutral 
current coupling of the b-quark. The helicity amplitudes are calculated
from resonant and background diagrams and the spin density matrix of the
final state is discussed. The spin analyzer of the $\tau$'s is illustrated
with the decays $\pi \nu$ and $\rho \nu \rightarrow (\pi \pi) \nu$. With
$10^8 \Upsilon$ a sensitivity to $g_V^b$ of a few per cent could be 
reachable. 

\end{abstract}
\end{center}
\end{titlepage}
\newpage
\pagestyle{plain}
\pagenumbering{arabic}
\vspace{4cm}

\section{Introduction}

A precise determination of the fermionic electroweak couplings can provide 
stronger hints on the nature of new physics at higher scales through
the quantum corrections to the effective theory at lower energies. 
At present there is no solid experimental discrepancy with the predictions
of the Standard Model \cite{pdg}, and this fact is used to set bounds on these
new dynamics beyond the Standard Model.

In this context, the agreement between the measured $Z- f\bar{f}$ couplings
and the Standard Model is very good. However, while the lepton couplings
have been independently measured for the three families at LEP, for the
quarks only the $b$ and $c$ events can be separated in the hadronic event
sample and consequently their couplings measured exclusively. From the 
measurements of $R_b$ and $A_b$ \cite{eww} we can get the values of the
vector and axial couplings of Z to $b\bar{b}$ \cite{schaile}. To find the 
accuracy on the bottom-Z couplings obtainable from these measurements it is
enough to use tree level expressions for $R_b$ and $A_b$ because higher order
corrections will slightly shift the central values but will not modify the
errors. Using the measurements quoted in \cite{eww}, and defining the
vector and axial couplings in the Standard Model as $g_V^b = -1/4 + 
\sin^2 \theta_W/3$ and $g_A^b=1/4$, the accuracy on the
bottom-Z couplings is: $\delta g_V^b =\pm 0.013$ and 
$\delta g_A^b =\pm 0.007$. It is important to notice that while the 
uncertainty in the $g_A^b$ coupling is of $2.8 \%$, the $g_V^b$ 
measurent is worse, with an uncertainty of $7.5 \%$.  

For the light quarks this separation has not been achieved at LEP and so an
exclusive determination of their couplings is not available. A study of the
final state distributions at different meson facilities can provide
independent measurements of these couplings. 
In a previous work \cite{phi}, we showed
that a $\Phi$ factory with polarized $e^-$ beams can supply the information
on the s-quark couplings.

We will show in this work, that a detailed study of the final state
distributions of the decay products of the $\tau$'s for $e^+e^- \rightarrow
\Upsilon \rightarrow \tau^+\tau^-$ would provide valuable information on the 
vector $Z-b\bar{b}$ coupling, $g_V^b$.  

To determine this coupling at energies well below the Z pole, we will have 
to measure the interference between $\gamma$ and Z, where Z is coupled 
to a $b\bar{b}$ resonance.
The $b \bar{b}$ mesons which can couple both to $\gamma$ and Z are the
$\Upsilon$ mesons. We are interested in a process in which the $\Upsilon$
is coupled to a Z either in its production or in its decay. This means we 
will study the decay of a polarized $\Upsilon$ or its weak decay.

For these reasons, we will analyze the leptonic decays of the $\Upsilon$ 
resonances, therefore we could use $\Upsilon(1S)$, $\Upsilon(2S)$ and 
$\Upsilon(3S)$, but 
not $\Upsilon(4S)$, that decays dominantly to $B\bar{B}$ where the 
information on the $\Upsilon$ polarization is lost. In this work
we will concentrate on the $\Upsilon(1S)$ resonance, but everything would 
be similar for $\Upsilon(2S)$ and $\Upsilon(3S)$. 
We can see in Table 1 that the branching ratios of $\Upsilon(1S)$ to the 
three charged 
leptons are approximately $3 \%$, but $e^+e^-$ and $\mu^+ \mu^-$ can not be 
used because their polarizations are not measurable in the detector 
through decay distributions. Obviously $e^+ e^-$ are stable particles and
$\mu^+ \mu^-$ at this 
energy do not decay inside the detector. As we will see explicitely later,
all the relevant information on $g_V^b$, that comes from a P-odd $\gamma-Z$
interference, only appears at leading order in the polarization of the 
final leptons. This means that we are constrained to consider the decay 
$\Upsilon \rightarrow \tau^+ \tau^-$ and to measure the $\tau$ polarizations.

Before entering a complete analysis of the $\tau$ observables, to estimate 
the sensitivity of this process to the vector coupling, let us 
first consider the $\tau^-$ longitudinal polarization, suggested in Ref. 
\cite{pascual} in another context. We will make a reasonable approximation 
in order to get
a simple and clean result: the resonant diagrams will dominate the 
process on the
$\Upsilon(1S)$ peak. So, we consider only diagrams (2), (3) and (4) in Fig. 1.
Under these conditions we get a longitudinal polarization from the parity
violating interferences of the dominant amplitude (2) with the neutral
current amplitudes (3) + (4).  

\beq
\label{justpol}
P_{z'}\ =\ 2\ \frac{8 G_F}{\sqrt{2}}\ \frac{s}{4 \pi \alpha}\ g_A\ 
\frac{g_V^b}{Q_b}\ \frac{(1+ \cos^2\theta)|\vec{p}|\ p^0  + 2 
\cos \theta (p^0)^2 }{(1 + \cos^2 \theta) (p^0)^2  + \sin^2 \theta M_\tau^2}
\eeq
where $g_A=1/4$ is the axial coupling to the leptons, $Q_b$ the charge 
of the b quark and $p^\mu = (p^0, \vec{p}(\theta))$ the four-momentum 
of the $\tau^-$.

In this expression we can discover some interesting features of this 
observable,
\begin{itemize}
\item It is linear on $g_V^b$, the vector coupling which we want to 
determine, showing up together with the axial coupling of the Z to leptons.
\item The magnitude of the $\tau$-polarization on the $\Upsilon$ peak is 
set by the factor $8 G_F M^2_\Upsilon / ( 4 \pi \alpha \sqrt{2} ) \simeq 
0.064$, that translates into $P_{z'}(\theta =0) = 0.032$.
\item It is independent of the hadronic structure of the 
resonance which cancels completely in the ratio.
\end{itemize}

All these properties will be modified when we include the
non-resonant diagrams, but these new contributions will correct this result
at the level of a few per cent, so this will be the dominant behaviour of 
our observables.
These properties were already pointed out in Ref. \cite{nucl}, where 
it was shown that this polarization is enhanced in the 
vicinity of the resonance. Approximately at four amplitudes below the 
resonance one gets a polarization five times bigger, but the number of 
events decreases three orders of magnitude and so it is 
more efficient from the point of view of statistics to stay on the peak 
of the resonance. 

The above example indicates that a more detailed analysis 
of the problem is worth. In the next section we calculate 
the complete $\tau^-\tau^+$ density 
matrix in this process which contains all the relevant information on 
$g_V^b$. In section 3 we study the hadronic decays of the $\tau$ 
to meassure the $\tau$ density matrix, and from here we analyse, in 
section 4, the statistical accuracy that is possible to get in the 
meassure of the vector coupling to the b quark in each channel.
 
\section{$\tau^+\tau^-$ density matrix}

The density matrix from $e^-(l_-,\xi_-) 
e^+(l_+,\xi_+)\rightarrow \tau^-(p_-,\lambda_-) \tau^+(p_+,\lambda_+)$,
in terms of helicity amplitudes \cite{martin}, and when our initial
beams are unpolarized, Eq. (\ref{2density}), is given by, 

\begin{equation}
\label{densid-P}
\rho^{\tau}_{(\lambda_-,\lambda_+),(\lambda_-^\prime,\lambda_+^\prime)}
=\sum_{\vec{\xi}} f_{(\lambda_-,\lambda_+),\vec{\xi}}
\ (\theta)    f^*_{(\lambda_-^\prime,\lambda_+^\prime),\vec{\xi}}
\ (\theta)
\end{equation}
where the angle $(\theta)$ is given by the direction of the $\tau^-$
relative to the initial $e^-$ beam, with the x-z plane defined as the 
scattering plane.

Using reduced helicity amplitudes, $T^J_{\vec{\lambda}\xi}$, and 
rotation matrices these helicity amplitudes are \cite{martin},
\beq
\label{ampli 2}
f_{\vec{\lambda},\vec{\xi}}\ (\theta) = 
d^{J}_{\xi,\lambda}(\theta)\, T^J_{\vec{\lambda}\xi}
\eeq
where $\lambda=\lambda_--\lambda_+$, $\xi=\xi_--\xi_+$ and 
$d^{J}_{\xi,\lambda}(\theta)$ are the reduced rotation matrices about the
y-axis.
If we neglect the electron mass, the total angular momentum of the 
process is always $J=1$. Therefore we get Eq. (\ref{ampli 2}) where we have
re-defined our reduced helicity amplitudes with
respect to Ref. \cite{martin}, including in our definition several
normalization factors irrelevant in our analysis \cite{phi}. 
Furthermore, helicity conservation in the electron vertex implies that
$\xi$ fixes completely $\vec{\xi}=(\xi_+, \xi_-)$.
These reduced helicity amplitudes get contributions from diagrams (1) to (5)
in Figure 1.
For the dominant amplitudes contributing through interferences to P-odd
and C-odd observables we can write

\begin{eqnarray}
\label{resonan}
T_{\vec{\lambda},\xi}\,=\,T_{\vec{\lambda},\xi}(\gamma)\,+
\, T_{\vec{\lambda},\xi}(\gamma Z_A)\,
+\, T_{\vec{\lambda},\xi}(Z_A \gamma)\,+\,T_{\vec{\lambda},\xi}(Z_{A,A})  
\end{eqnarray}
where $T_{\vec{\lambda},\xi}(\gamma)$ accounts for the contribution of 
diagrams (1) and (2).  
$T_{\vec{\lambda},\xi}(\gamma Z_A)$ is the P-violating
piece of diagram (3) plus the $VA$ piece of diagram (5) and $T_{\vec{\lambda},
\xi}(Z \gamma_A)$ the corresponding P-violating piece of diagram (4) plus the
$A V$ piece of diagram (5). Finally $T_{\vec{\lambda},\xi}(Z_{A,A})$ is 
the contribution of diagram (5) with axial couplings in the initial and
final vertices. Notice that here we have not included the 
$T_{\vec{\lambda},\xi}(\gamma Z_V)$, $T_{\vec{\lambda},\xi}(Z \gamma_V)$ and
$T_{\vec{\lambda},\xi}(Z_{V,V})$ pieces, because they are sub-dominant with
respect to $T_{\vec{\lambda},\xi}(\gamma)$, both in the resonant and 
non-resonant components.

Taking into account the transformation properties under P of these amplitudes
we get

\begin{eqnarray}
\label{parity}
T_{\vec{\lambda},\xi}(\gamma)  = &  T_{-\vec{\lambda},-\xi}(\gamma)\ \  = &    
T_{\vec{\lambda},-\xi}(\gamma) \ \ \ =   T_{-\vec{\lambda},\xi}(\gamma) \\
T_{\vec{\lambda},\xi}(\gamma Z_A) = &- T_{-\vec{\lambda},-\xi}(\gamma Z_A)  
= & T_{\vec{\lambda},-\xi}(\gamma Z_A) =  - T_{-\vec{\lambda},\xi}(\gamma Z_A)
 \nn \\
T_{\vec{\lambda},\xi}(Z_A \gamma) = &- T_{-\vec{\lambda},-\xi}(Z_A \gamma)  
= & - T_{\vec{\lambda},-\xi}(Z_A \gamma)  =  T_{-\vec{\lambda},\xi}
(Z_A \gamma) \nn \\
T_{\vec{\lambda},\xi}(Z_{AA})  = &  T_{-\vec{\lambda},-\xi}(Z_{AA})  = &   
-T_{\vec{\lambda},-\xi}(Z_{AA})  =  -T_{-\vec{\lambda},\xi}(Z_{AA}) \nn 
\end{eqnarray}
where the normalization of these reduced helicity amplitudes is such that
\begin{eqnarray}
\label{cross}
\frac{d \sigma}{d \Omega}&=& Tr(\rho^{out}) = 2 \sin^2\theta |T_{(+,+),1}
(\gamma)|^2 + (1 +\cos^2 \theta) |T_{(+,-),1}(\gamma)|^2 \nn \\
&+& 4 \cos \theta Re\{T_{(+,-),1}(\gamma) T^*_{(+,-),1}(Z_{A,A}) \}
\end{eqnarray}

\beq
\label{totcross}
\sigma = \frac {16 \pi}{3} (|T_{(+,+),1}(\gamma)|^2 + |T_{(+,-),1}(\gamma)|^2)
\eeq
Notice that a C-odd forward-backward interference is generated by the
axial couplings of the Z in both the electron and $\tau$ vertices.

We can calculate these reduced helicity amplitudes from the Feynman diagrams
1-5 of Fig. 1 following the method explained in App. B.

\begin{eqnarray}
\label{amplred}
K T_{(+,+),1}(\gamma)&=& -i 4 \sqrt{2}\ \frac{e^2}{s}\ (1 +\frac{e^2}{s} Q_b^2 
|F_\Upsilon|^2 P_\Upsilon)\ M_\tau\ \frac{\sqrt{s}}{2}\nn \\
K T_{(+,-),1}(\gamma)&=& -i 8\ \frac{e^2}{s}\ (1 + \frac{e^2}{s} Q_b^2 
|F_\Upsilon|^2 P_\Upsilon)\ p^0\ \frac{\sqrt{s}}{2} \nn \\
K T_{(+,+),1}(\gamma Z_A)&=& 0 \nn \\
K T_{(+,-),1}(\gamma Z_A)&=& -i 8\ \frac{8 G_F}{\sqrt{2}}\ g_A\ (\frac{e^2}{s} 
Q_b g_V^b |F_\Upsilon|^2 P_\Upsilon-g_V)\ |\vec{p}|\ \frac{\sqrt{s}}{2} \nn \\
K T_{(+,+),1}(Z_A \gamma)&=&  -i 4 \sqrt{2}\ \frac{8 G_F}{\sqrt{2}}\ g_A\ 
(\frac{e^2}{s} Q_b g_V^b |F_\Upsilon|^2 P_\Upsilon-g_V)\ M_\tau\  
\frac{\sqrt{s}}{2}\nn \\
K T_{(+,-),1}(Z_A \gamma)&=& -i 8\ \frac{8 G_F}{\sqrt{2}}\ g_A\ (\frac{e^2}{s} 
Q_b g_V^b |F_\Upsilon|^2 P_\Upsilon-g_V)\ p^0\ \frac{\sqrt{s}}{2}\nn \\
K T_{(+,+),1}(Z_{AA})&=& 0   \nn \\
K T_{(+,-),1}(Z_{AA})&=& i 8\ \frac{8 G_F}{\sqrt{2}}\ g_A^2\ |\vec{p}|\ 
\frac{\sqrt{s}}{2}            
\end{eqnarray}
where K is only a constant which takes care of the different normalization
of the helicity amplitudes and the Feynman amplitudes. $Q_b = -\frac{1}{3}$
is the charge of the b quark, $g_{V(A)}$ is the vector (axial) coupling to 
the leptons, $P_\Upsilon$ stands for the Breit-Wigner of the $\Upsilon$,
\beq
P_\Upsilon (s) = \frac{1}{(s-M_\Upsilon^2) + i M_\Upsilon \Gamma_\Upsilon}
\eeq
and $F_\Upsilon (q^2)$ is a form factor defined as,
\beq
\langle \Upsilon (q,\omega) | \bar{\psi}_b(0) \gamma_\mu \psi_b(0) 
| 0 \rangle = F_\Upsilon (q^2) \varepsilon^*(\omega,q)_\mu
\eeq
with $\varepsilon^*(\omega,q)_\mu$, the polarization four-vector.
This form factor can be related to the amplitude for $\Upsilon$ to $e^+ e^-$,
\beq
\Gamma_e = \frac{1}{6\pi} Q_b^2 \frac{(4 \pi \alpha)^2}{M_\Upsilon^4}
|F_\Upsilon|^2 \frac{M_\Upsilon}{2}
\eeq
Notice that all the hadronic uncertainties in our process will be 
included in this unique form factor.
In Eq. (\ref{amplred}) we can see the coupling $g_V^b$ is only in the 
$T_{\vec{\lambda},\xi}(\gamma Z_A)$ and $T_{\vec{\lambda},\xi}(Z_A \gamma)$ 
amplitudes, so, as these amplitudes will contribute to dominant order
to the observables through interferences with $T_{\vec{\lambda},\xi}
(\gamma)$, this means that only the P-odd observables will contain the 
information about the $g_V^b$ coupling linearly. Then we are going to 
analyze the polarizations and the P-odd correlations.

The polarizations of $\tau^-$ are given as follows
\begin{eqnarray}
\label{polzm}
\frac{d \sigma}{d \Omega} P_{z'}^{(-)}(\theta) &=&   \rho_{(+,+),(+,+)} + 
\rho_{(+,-),(+,-)} - \rho_{(-,+),(-,+)} - \rho_{(-,-),(-,-)} \nn\\
&=& 2 Re \{T_{(+,-),1}(\gamma)T^*_{(+,-),1}(\gamma Z_A)\} (1 + \cos^2 \theta) \nonumber
\\& + & 4 Re \{T_{(+,-),1}(\gamma)T^*_{(+,-),1}(Z_A \gamma)\} \cos \theta 
\end{eqnarray}

\begin{eqnarray}
\label{polxm}
\frac{d \sigma}{d \Omega}P^{(-)}_{x'}(\theta) &=&  \rho_{(+,+),(-,+)} + 
\rho_{(+,-),(-,-)} + \rho_{(-,+),(+,+)} + \rho_{(-,-),(+,-)}\nn \\
&=& -2 \sqrt{2} [  
Re \{T_{(+,+),1}(\gamma )T^*_{(+,-),1}(\gamma Z_A)\} \sin \theta\ 
\cos \theta \nonumber \\
& + & ( Re \{T_{(+,+),1}(\gamma )T^*_{(+,-),1}(Z_A \gamma)\} \nonumber \\
& + & Re \{T_{(+,-),1}(\gamma )T^*_{(+,+),1}(Z_A \gamma)\}) \sin \theta ] 
\end{eqnarray}

\begin{eqnarray}
\label{polym}
\frac{d \sigma}{d \Omega} P^{(-)}_{y'}(\theta) &=&-i( \rho_{(+,+),(-,+)} + 
\rho_{(+,-),(-,-)} - \rho_{(-,+),(+,+)} - \rho_{(-,-),(+,-)})\nn \\
&=&  2 \sqrt{2}   
[Im \{T_{(+,+),1}(\gamma ) T^*_{(+,-),1}(Z_{A,A})\} \sin \theta \nn \\
&+& 2 Im \{T_{(+,+),1}(\gamma ) T^*_{(+,-),1}(\gamma)\} \sin \theta \cos 
\theta] 
\end{eqnarray}
As we can see in these expressions, only $P_{z'}$ and $P_{x'}$
contain information on $g_V^b$, because both are P-odd, T-even observables,
but $P_{y'}$ has no information, because it is P-even, T-odd.
The T-odd observable $P_{y'}$ needs the interference between resonant and 
non-resonant amplitudes, that on the $\Upsilon$ peak are relatively 
imaginary. By the same argument, there is no interference between
resonant and non-resonant pieces in the T-even observables, like $P_{z'}$ 
and $P_{x'}$, on the $\Upsilon$ peak. 
If we compare the longitudinal polarization, Eq. (\ref{polzm}), with the 
result we obtained in Eq. (\ref{justpol}) the difference is in
the non-resonant terms proportional to $g_V$ in Eq. (\ref{amplred}). 
Then these new terms are suppressed on the $\Upsilon$ peak by 
a factor $g_V \alpha^2 Q_b /(9 g_V^b b.r.(\Upsilon \rightarrow e^+ e^-)^2) 
\approx 3.5 \times 10^{-4}$, and so our estimate in Eq. (\ref{justpol}) 
is very good.
This means basically, that this observable will be, in the forward 
direction, $P_{z'}(\theta=0) \simeq .185\cdot g_V^b \simeq 0.032$. In Fig.
2, we can see a plot of this polarization as a function of $\theta$.

On the other hand $P_{x'}$ contains similar reduced helicity amplitudes
as $P_{z'}$, the only difference is that, as usual, transverse polarizations
are suppressed by a mass insertion, that is a factor $M_\tau/p^0 \simeq .38$. 
Basically, apart from a different angular dependence, this is the reason
that makes this observable less sensitive to $g_V^b$ as we can see
in Fig. 3. We get for instance $P_{x'}(\theta=\pi/2) \simeq
- .063 \cdot g_V^b$, slightly worse than $P_{z'}$.
  
The information contained in the $\tau^+$ polarizations is closely 
related to that of the $\tau^-$
\begin{eqnarray} 
P_{z'}^{(+)} = - P_{z'}^{(-)} \nn \\
P_{x'}^{(+)} = P_{x'}^{(-)} \nn \\
P_{y'}^{(+)} = - P_{y'}^{(-)}
\end{eqnarray}

Finally, we can also consider the spin correlations between $\tau^+$
and $\tau^-$. We define these spin correlations as follows,

\begin{eqnarray}
\label{correlations}
\frac{d \sigma}{d \Omega} {\cal C}_{i j}(\theta) &=& Tr(\sigma_i^{(-)} \sigma_j
^{(+)} \rho^\tau) 
\end{eqnarray}
With this definition is easy to see that the only P-odd observables will
be the ${\cal C}_{x,y}$, ${\cal C}_{z,y}$, ${\cal C}_{y,x}$ and ${\cal C}_{y,z}$ correlations,

\begin{eqnarray}
\label{cozy}
\frac{d \sigma}{d \Omega}{\cal C}_{zy}(\theta) &=& i (- \rho_{(+,+),(+,-)} + 
\rho_{(-,+),(-,-)} + \rho_{(+,-),(+,+)} - \rho_{(-,-),(-,+)}) \nn\\
&=& -2 \sqrt{2} [  
Im \{T_{(+,+),1}(\gamma )T^*_{(+,-),1}(\gamma Z_A)\}
\sin \theta\ \cos \theta \nonumber
\\& + & ( Im \{T_{(+,+),1}(\gamma )T^*_{(+,-),1}(Z_A \gamma)\} \nonumber \\
&-& Im \{T_{(+,-),1}(\gamma )T^*_{(+,+),1}(Z_A \gamma)\}) \sin \theta ] 
\end{eqnarray}

\begin{eqnarray}
\label{coyz}
{\cal C}_{yz}(\theta) &=& {\cal C}_{zy}
(\theta)
\end{eqnarray}

\begin{eqnarray}
\label{coxy}
\frac{d \sigma}{d \Omega} {\cal C}_{xy}(\theta) &=& i ( -\rho_{(+,+),(-,-)} - 
\rho_{(-,+),(+,-)} + \rho_{(+,-),(-,+)} + \rho_{(-,-),(+,+)}) \nn \\
&=&  2 Im\{T_{(+,-),1}(\gamma)T^*_{(+,-),1}(\gamma Z_A)\} \sin^2\theta
\end{eqnarray}

\begin{eqnarray}
\label{coyx}
{\cal C}_{yx}(\theta) &=& - {\cal C}_{xy}(\theta)
\end{eqnarray}
As we already pointed out and we can see explicitly in Eq. 
(\ref{polzm})-(\ref{coyx}),
all the relevant observables are associated to the amplitudes 
$T_{\vec{\lambda},\xi}(\gamma Z_A)$ and $T_{\vec{\lambda},\xi}(Z_A \gamma)$,
and the relative strength of these observables to the dominant term is set
by the factor $8 G_F M^2_\Upsilon / ( 4 \pi \alpha \sqrt{2} ) \simeq 0.064$.
Unfortunately, the P-odd correlations are also T-odd which means
they need an imaginary part. With our set of reduced helicity amplitudes, 
it becomes 
necessary to have an interference between a resonant and a non-resonant
diagram to get an imaginary part. Therefore, these contributions have the 
same suppression that diagram (1) with respect to diagram (2), that is,
$\alpha/(3\ b.r.(\Upsilon \rightarrow e^+e^-))\simeq 1/10$. 
Notice also that the helicity structure of ${\cal C}_{zy}$ is very 
similar to $P_{x'}$, and so it will have the same suppression factors, 
unlike ${\cal C}_{xy}$ which is not helicity suppressed. 

Then it is clear that our main observables will be the longitudinal 
polarizations of both $\tau$'s, then $P_{x'}$ and finally ${\cal C}_{xy}$
and ${\cal C}_{zy}$ ordered from the most relevant to the least one.

In the next section we are going to connect these observables to measurable
quantities, analyzing the 
angular distributions of the decay products of the two $\tau$.

\section{Decay of a polarized $\tau$}

The main $\tau$ decay channels are presented in Table 2. The purely leptonic
decays have branching ratios of $17.65\%$ for muons and $18.01\%$ for 
electrons. Unfortunately, these decay modes have two neutrinos in the final
state, which implies we can not reconstruct the $\tau$ direction. Then, 
their sensitivity to the $\tau$ polarization is small 
\cite{mart-hagi} compared with the hadronic decays. 

So we will concentrate on the hadronic $\tau$ decays which have only one 
neutrino in the final state, which
allows to reconstruct the $\tau$ direction if both $\tau$ decay hadronically
\cite{kuhn-dir}. These decays are $\tau^- \rightarrow \pi^- \nu_\tau$, with a 
branching ratio of $11.31\%$, $\tau^- \rightarrow \rho^- \nu_\tau$ which corresponds 
almost exactly to the two pions channel, with branching ratio $25.24\%$,
and $\tau^- \rightarrow a_1^- \nu_\tau$ which is given by the sum of the three
pion final states. In this work we will concentrate, as an example, in the 
decays $\tau^-\rightarrow \pi^- \nu_\tau$ and $\tau^-\rightarrow \rho^- 
\nu_\tau$, other $\tau$ decay channels have been studied elsewhere, for
instance $\tau \rightarrow a_1 \nu_\tau$ can be found in \cite{rouge} for
LEP physics,

\subsection{$\tau^- \rightarrow \pi^- \nu_\tau$}

This channel has been used for a long time to measure the $\tau$ 
polarization because of its good sensitivity. In this decay, we can easily 
get the differential decay width as,

\begin{equation}
\label{pidis}
\frac{1}{\Gamma} \frac{d \Gamma}{d \Omega} = \frac{1}{4 \pi} [ 1 +
\vec{P} \hat{k}(\Omega)]
\end{equation}
where $\hat{k}$ is the unit vector in the direction of the pion.

In Eq. (\ref{pidis}) we can see that the polarization effects are
not suppressed in the angular distribution of the decay pions.
This is due to the fact that in this process there is only one reduced 
helicity amplitude, and so the complete angular distribution is
necessarily proportional to this unique reduced amplitude.
So this channel is specially sensitive to
the $\tau$ polarization, and this has been the reason for its
popularity as polarization analyzer in $\tau$ decays.

With these elements, following App. A, we can build a complete two steps 
angular distribution \cite{phi,nelson} for the whole process,
$e^-e^+\rightarrow \tau^- \tau^+\rightarrow (\pi^- \nu_\tau)  (\pi^+ \bar{\nu}_\tau)$

\begin{eqnarray}
\label{final}
&&\frac{d \sigma}{d \Omega\ d \Omega_+\ d \Omega_-} = \frac{d \sigma}{d \Omega}\ ( 1\ +\ \vec{P}^{(-)}\cdot 
\hat{k}_-\ -\ \vec{P}^{(+)}\cdot \hat{k}_+ \nn \\
&&- {\cal C}_{zz} \cos \theta_-\cos \theta_+\ -\ {\cal C}_{xx} \sin \theta_- \cos \phi_- 
\sin \theta_+ \cos \phi_+ \nn \\ 
&&- {\cal C}_{yy} \sin \theta_-\sin \phi_- \sin \theta_+ \sin \phi_+ \ -\ 
{\cal C}_{zx} \cos \theta_- \sin \theta_+ \cos \phi_+ \nn \\
&&- {\cal C}_{xz} \sin \theta_- \cos \phi_- \cos \theta_+ \ -\ {\cal C}_{zy} \cos \theta_-
\sin \theta_+ \sin \phi_+ \nn \\
&&- {\cal C}_{yz} \sin \theta_- \sin \phi_-\cos \theta_+\ -\ {\cal C}_{xy} \sin \theta_- 
\cos \phi_- \sin \theta_+ \sin \phi_+ \nn \\ 
&&- {\cal C}_{yx} \sin \theta_- \sin \phi_- \sin \theta_+ \cos \phi_+)
\end{eqnarray} 
where $(\theta_\pm, \phi_\pm)$ is the direction of the $\pi^\pm$ in the rest
frame of the $\tau^\pm$ and $\hat{k}_\pm$ is the unit vector in this 
direction.

This will be the cross section we have to study to extract $g_V^b$ from 
this channel. In section 4 we will analyze the sensitivity of the 
different observables we can construct from this cross section.

\subsection{ $\tau^- \rightarrow \rho^- \nu_\tau \rightarrow \pi^- \pi^0
\nu_\tau$}

In the channel $\tau \rightarrow \rho \nu_\tau$ we have a spin 1 particle in
the final state. This implies that we have two different helicity
amplitudes in this decay, and so different combinations of these amplitudes
enter in the polarized and unpolarized pieces,

\begin{equation}
\frac{1}{\Gamma_{\tau\rightarrow\rho\nu}} \frac{d\Gamma_{\tau\rightarrow\rho\nu}}{d \Omega} =
 \frac{1}{4 \pi} [ 1 +\alpha_\rho \vec{P} \hat{k}(\Omega)]
\end{equation}
where $\alpha_\rho$ is a ratio of reduced helicity amplitudes, which 
we can get in terms of the masses as,

\begin{equation}
\label{alr}
\alpha_\rho = \frac{|T_{0,-1/2}|^2 - |T_{-1,-1/2}|^2}{|T_{0,-1/2}|^2 + 
|T_{-1,-1/2}|^2}=\frac{M_\tau^2 - 2 M_\rho^2}{M_\tau^2 + 2 M_\rho^2}= 0.456
\end{equation}

Then, we can see from here that in spite of its bigger statistics the 
sensitivity at this level is smaller than in $\tau \rightarrow \pi 
\nu_\tau$.
However, as it was pointed out in \cite{mart-hagi}, this situation can be 
improved if, in addition, we try to get some extra information on the
$\rho$ helicity. To do this, you have to include another 
step in this chain of decays, and analyze the decay $\rho^- \rightarrow
\pi^- \pi^0$. The cross section for the whole process, 
$e^-e^+\rightarrow \tau^- \tau^+\rightarrow ((\pi^-\pi^0) \nu_\tau)  
(\rho^+ \bar{\nu}_\tau)$, can be written as,

\begin{eqnarray}
\label{3steps}
\frac{d\sigma}{d\Omega d\Omega_1^- d\Omega_1^+ d\Omega_2}= 
\frac{d\sigma}{d\Omega}
(e^+ e^- \rightarrow \tau^+ \tau^-)\ \frac{1}{\Gamma_\tau^2}
\frac{d\Gamma_{\tau^- \tau^+\rightarrow \rho^- \nu \rho^+\bar{\nu}}}
{d\Omega_1^- d\Omega_1^+} \nonumber \\
\frac{1}{\Gamma_\rho}
\frac{d\Gamma_{\rho^- \rightarrow \pi^- \pi^0}}{d\Omega_2}
\end{eqnarray}  
where the $\tau^- \tau^+$ decay amplitude to $\rho^- \nu \rho^+ \bar{\nu}$ is
given below in Eq. (\ref{tworho}), and we have added the last term which 
is the decay width of a 
polarized and aligned $\rho$ into two pions. The expression for this 
decay width is

\beq
\label{rhodec}
\frac{1}{\Gamma_{\rho\rightarrow\pi\pi}}\frac{d\Gamma_{\rho \rightarrow \pi \pi}}
{d\Omega_2}= \frac{1}{4\pi}
\, [1 -\sqrt{10} \sum_N {\cal D}^{(2)}_{N,0}(\phi_2,\theta_2,0) t_{2,N}]
\eeq
In this equation we can see that the $\rho$ polarizations do not appear in
the decay angular distribution because this is a strong decay and
therefore P-conserving. The alignments, higher order multipole 
parameters, appear but the polarizations do not.

Now, we have to calculate the density matrix for a $\rho$ coming from the 
decay of a polarized $\tau$ in its center of mass frame, and then apply 
the necessary Wigner rotation \cite{wigner}, as explained in App. A,
to get the density matrix in a frame where the $\tau$ is moving.

This density matrix of a single $\rho$ from the decay of one of the
two $\tau$s also contains information on the decay of the other
$\tau$ if we study the correlations and do not integrate the second
$\tau$ decay. So, as we are interested in the measure of correlations between
the two $\tau$s, we will study the $\rho^-$ density matrix in a
decay $\tau^-\tau^+ \rightarrow (\rho^- \nu_\tau) (\rho^+ \bar{\nu}_\tau)$
and in a decay  $\tau^-\tau^+ \rightarrow (\rho^- \nu_\tau) (\pi^+ \bar{\nu}_\tau)$.
Naturally this two different density matrices will coincide when we 
integrate completely the $\tau^+$ decay products.

Following Appendix A, we can write the complete $\rho^-$ density matrix
from a decay $\tau^-\tau^+ \rightarrow (\rho^- \nu_\tau) (\rho^+ \bar{\nu}_\tau)$
as

\begin{eqnarray}
\label{rhodensity}
\rho_{\mu_- \mu_-^\prime}&=\sum_{\lambda_+\lambda_-\lambda_+^\prime\lambda_-^\prime}
\sum_{\nu_- \nu_-^\prime \nu_+} f_{(\nu_+ 1/2) \lambda_+}^{(+)}(\Omega_+)
d_{\mu_- \nu_-}(\omega_-) f_{(\nu_- -1/2) \lambda_-}^{(-)}(\Omega_-) \nn \\
&\rho^{\tau}_{(\lambda_-,\lambda_+),(\lambda_-^\prime,\lambda_+^\prime)}
f_{(\nu_+ 1/2) \lambda_+^\prime}^{(+) *}(\Omega_+)d_{\mu_-^\prime \nu_-^\prime}
(\omega_-) f_{(\nu_-^\prime -1/2) \lambda_-^\prime}^{(-) *}(\Omega_-)
\end{eqnarray}

where we have used a reference system in LAB with the $\tau^-$ in the z-axis
and the initial beams in the x-z plane, to simplify the expressions for
the Wigner rotations.

We define an effective density matrix, $\bar{\rho}$ without the Wigner 
rotations that, if we integrated completely the $\Omega_+$ variables,
would correspond to the density matrix in the $\tau^-$ rest frame, 
which is 
\beq
\label{effdensity}
\rho_{\mu_- \mu_-^\prime}=d_{\mu_- \nu_-}(\omega_-)\bar{\rho}_{\nu_- 
\nu_-^\prime}d_{\mu_-^\prime \nu_-^\prime}(\omega_-)
\eeq

The next step is to calculate this effective density matrix in terms of 
reduced helicity amplitudes. In a decay we define the helicity amplitudes as,
\beq
\label{ampldec}
f_{\vec{\nu} \lambda}(\theta,\phi) =\sqrt{\frac{2j+1}{4\pi}} 
{\cal D}^{j *}_{\lambda, \nu_1-\nu_2}(\phi,\theta,0) T^j_{\vec{\nu}} 
\eeq
With this definition and Eqs. (\ref{rhodensity}) and (\ref{effdensity}), 
we get our effective density matrix.
In this process, $\tau^- \rightarrow \rho^- \nu_\tau$,
we have only two reduced helicity amplitudes if we take $M_\nu=0$
\begin{eqnarray}
\label{aredrho}
K T_{-1,-1/2}&=& i 4 G_F\ V_{ud}\ F_\rho(q^2)\ \sqrt{M_\tau k_2}\nn \\
K T_{0,-1/2} &=& i 2\sqrt{2}G_F\ V_{ud}\ F_\rho(q^2)\ \sqrt{M_\tau k_2} 
\ \frac{M_\tau}{M_\rho} 
\end{eqnarray}
and also two amplitudes in the decay $\tau^+ \rightarrow \rho^+ \bar{\nu}_\tau$, 
\begin{eqnarray}
\label{aredrho2}
K T_{1,1/2}&=& -i 4 G_F\ V_{ud}\ F_\rho(q^2)\ \sqrt{M_\tau k_2}\nn \\
K T_{0,1/2} &=& i 2\sqrt{2}G_F\ V_{ud}\ F_\rho(q^2)\ \sqrt{M_\tau k_2} 
\ \frac{M_\tau}{M_\rho} 
\end{eqnarray}
where $F_\rho (q^2)$ is a form factor defined as,
\beq
\langle \rho (k,\omega) |T| \pi (\vec{k_2}) \pi(-\vec{k_2}) \rangle = 
F_\rho (q^2) \varepsilon^*(\omega,q)_\mu k^\mu_2
\eeq
The rest of the notation is self-explanatory.
The total amplitude $\rho \rightarrow \pi \pi$ is,
\beq
\Gamma_{\rho\rightarrow\pi\pi}= \frac{1}{2 M_\rho} (|T_{-1,-1/2}|^2 + |T_{0,-1/2}|^2)
\eeq

Following App. A, we now calculate the C.M. multipole parameters
corresponding to the density matrix $\bar{\rho}$ in terms of these reduced 
amplitudes, taking into account that we do not integrate the direction of 
the second $\rho$. 

\begin{eqnarray}
\label{tworho}
&\frac{d\Gamma_{\tau^-\tau^+}}{d \Omega_1^- d \Omega_1^+}=
Tr\{ \rho \} = \rho_{-1,-1} + \rho_{0,0}\nn \\
& = (|T_{-1,-1/2}|^2 + |T_{0,-1/2}|^2)(|T_{1,1/2}|^2 + |T_{0,1/2}|^2) \nn \\
&[ 1 + \alpha_\rho \bar{P}\cdot \hat{k}^- - \bar{\alpha} 
\bar{P}\cdot \hat{k}^+ - \alpha_\rho \bar{\alpha} (  {\cal C}_{zz} \cos \theta_-\cos \theta_+\ \nn \\
&+\ {\cal C}_{xx} \sin \theta_- \cos \phi_- 
\sin \theta_+ \cos \phi_+ + {\cal C}_{yy} \sin \theta_-\sin \phi_- 
\sin \theta_+ \sin \phi_+ \nn \\ 
&+ {\cal C}_{zx} \cos \theta_- \sin \theta_+ \cos \phi_+ +
 {\cal C}_{xz} \sin \theta_- \cos \phi_- \cos \theta_+ \nn \\
& +\ {\cal C}_{zy} \cos \theta_-
\sin \theta_+ \sin \phi_+ + {\cal C}_{yz} \sin \theta_- \sin \phi_-
\cos \theta_+ \\
& +\ {\cal C}_{xy} \sin \theta_- \cos \phi_- \sin \theta_+ \sin \phi_+ 
+ {\cal C}_{yx} \sin \theta_- \sin \phi_- \sin \theta_+ \cos \phi_+)]
\nn
\end{eqnarray} 
here $(\theta_\pm, \phi_\pm)$ is the direction of the $\rho^\pm$ in the rest
frame of the $\tau^\pm$ and $\hat{k}_\pm$ is the unit vector in this 
direction. $\alpha_\rho$ was defined in Eq. (\ref{alr}) and $\bar{\alpha}$
is equal to  $\alpha_\rho$ if the $\tau^+$ decays to $\rho^+ \bar{\nu}_\tau$,
but is equal to 1 if it decays to $\pi^+ \bar{\nu}_\tau$. This definitions 
hold for all the observables that follow.

\begin{eqnarray}
&Tr\{ \rho \} \bar{t}_{2 0} = \sqrt{\frac{1}{10}}(\rho_{-1,-1} - 
2 \rho_{0,0})= \sqrt{\frac{1}{10}}(|T_{-1,-1/2}|^2 + |T_{0,-1/2}|^2)\nn \\
&(|T_{1,1/2}|^2 + |T_{0,1/2}|^2) \nn [\gamma_\rho -\beta_\rho \bar{P}^-\cdot 
\hat{k}^- - \bar{\alpha} \gamma_\rho \bar{P}^+\cdot \hat{k}^+ \nn \\
&+\bar{\alpha} \beta_\rho ({\cal C}_{zz} \cos \theta_-\cos \theta_+\ 
+\ {\cal C}_{xx} \sin \theta_- \cos \phi_- 
\sin \theta_+ \cos \phi_+ \nn \\
&+ {\cal C}_{yy} \sin \theta_-\sin \phi_- 
\sin \theta_+ \sin \phi_+  
+{\cal C}_{zx} \cos \theta_- \sin \theta_+ \cos \phi_+ \nn \\
&+ {\cal C}_{xz} \sin \theta_- \cos \phi_- \cos \theta_+ 
 +\ {\cal C}_{zy} \cos \theta_-
\sin \theta_+ \sin \phi_+\nn \\
&+ {\cal C}_{yz} \sin \theta_- \sin \phi_-
\cos \theta_+ +\ {\cal C}_{xy} \sin \theta_- \cos \phi_- \sin \theta_+ 
\sin \phi_+ \nn \\
&+ {\cal C}_{yx} \sin \theta_- \sin \phi_- \sin \theta_+ \cos \phi_+)]
\end{eqnarray}
where we have introduced two new coefficients $\beta_\rho$ and $\gamma_\rho$
defined as

\beq
\label{gamr}
\gamma_\rho = \frac{|T_{-1,-1/2}|^2 - 2 |T_{0,-1/2}|^2}{|T_{0,-1/2}|^2 + 
|T_{-1,-1/2}|^2}=\frac{2 M_\rho^2 - 2 M_\tau^2}{M_\tau^2 + 2 M_\rho^2}= -1.18
\eeq
\beq
\label{ber}
\beta_\rho  = \frac{|T_{-1,-1/2}|^2 + 2 |T_{0,-1/2}|^2}{|T_{0,-1/2}|^2 + 
|T_{-1,-1/2}|^2}=\frac{2 M_\rho^2 + 2 M_\tau^2}{M_\tau^2 + 2 M_\rho^2}= 1.73
\eeq
 
we also need $\bar{t}_{2 1}$ and $\bar{t}_{2 2}$,
\begin{eqnarray}
&Tr\{ \rho \} \bar{t}_{2 1} = \sqrt{\frac{3}{10}} \rho_{-1,0} 
= \sqrt{\frac{3}{10}}(|T_{-1,-1/2}|^2 + |T_{0,-1/2}|^2)
\nn \\
&(|T_{1,1/2}|^2 + |T_{0,1/2}|^2) [\delta_\rho 
( P_z^-(-\sin \theta_-) + P_x^- (\cos\theta_-\cos\phi_- 
- i\sin\phi_-) \nn \\
&+ P_y^- (\cos \theta_-\sin \phi_- + i\cos\phi_-))
-\delta_\rho \bar{\alpha} ({\cal C}_{zz}(-\sin \theta_-) \cos \theta_+
\nn \\
&+\ {\cal C}_{xx}(\cos\theta_-\cos\phi_- - i\sin\phi_-) 
\sin \theta_+ \cos \phi_+\nn \\
& + {\cal C}_{yy}(\cos \theta_-\sin \phi_- + 
i\cos\phi_-) \sin \theta_+ \sin \phi_+ + {\cal C}_{zx}(-\sin \theta_-) 
\sin \theta_+ \cos \phi_+ \nn \\
&+ {\cal C}_{xz}(\cos\theta_-\cos\phi_- - i\sin\phi_-) \cos \theta_+ 
 +\ {\cal C}_{zy}(-\sin \theta_-) \sin \theta_+ \sin \phi_+ \nn \\
&+ {\cal C}_{yz}(\cos \theta_-\sin \phi_- + 
i\cos\phi_-) \cos \theta_+ \nn \\
&+\ {\cal C}_{xy}(\cos\theta_-\cos\phi_- - 
i\sin\phi_-)\sin \theta_+ \sin \phi_+ \nn \\ 
&+ {\cal C}_{yx}(\cos \theta_-\sin \phi_- + i\cos\phi_-)\sin \theta_+ 
\cos \phi_+)]
\end{eqnarray}
\beq
Tr\{ \rho \} \bar{t}_{2 2} = \sqrt{\frac{3}{5}} \rho_{-1,1} = 0 
\eeq
we have also introduced a new coefficient
\beq
\label{delr}
\delta_\rho = \frac{T_{-1,-1/2} T_{0,-1/2}^*}{|T_{0,-1/2}|^2 + 
|T_{-1,-1/2}|^2}=\frac{\sqrt{2} M_\rho M_\tau}{M_\tau^2 + 2 M_\rho^2}= 0.445
\eeq
In general, these multipole parameters can be complex, as $\bar{t}_{2 1}$, 
except the parameters with $L=0$, which are always real.
  
It is very important to notice that unlike $\alpha_\rho$ and $\delta_\rho$,
that are small, both  $\gamma_\rho$ and  $\beta_\rho$ are bigger than one
and so they can enhance the information on the $\tau$ polarizations.

As we have already pointed out, the only difference between the decays  
$\tau^-\tau^+ \rightarrow (\rho^- \nu_\tau) (\rho^+ \bar{\nu}_\tau)$
and $\tau^-\tau^+ \rightarrow (\rho^- \nu_\tau) (\pi^+ \bar{\nu}_\tau)$ at this level 
is the value of the coefficient $\bar{\alpha}$

\begin{eqnarray}
\tau^+ \rightarrow \rho^+ \bar{\nu}_\tau \Longrightarrow \bar{\alpha}= .456  \nn \\
\tau^+ \rightarrow \pi^+ \bar{\nu}_\tau  \Longrightarrow \bar{\alpha}= 1  
\end{eqnarray}

The final step to get the alignments appearing in Eq. (\ref{rhodec}) and
Eq. (\ref{3steps}) is to make the Wigner rotation, Eq. (\ref{rotmulti}).

\begin{eqnarray}
&t_{2 0} = d^2_{0 M}(\omega_-) \bar{t}_{2 M} =\bar{t}_{2 0} (\frac{3}{2} 
\cos^2 \omega_- - \frac{1}{2}) \nn \\
&+ \sqrt{6} Re\{\bar{t}_{2 1}\} \sin \omega_- \cos \omega_-
\end{eqnarray}

\begin{eqnarray}
&t_{2 1} = d^2_{1 M}(\omega_-) \bar{t}_{2 M} = - \bar{t}_{2 0} \cos\omega_- 
\sin\omega_- + i Im\{\bar{t}_{2 1} \}\cos \omega_- \nn \\
&+ Re \{\bar{t}_{2 1} \} (\cos^2 \omega_- - \sin^2 \omega_-)
\end{eqnarray}

\begin{eqnarray}
&t_{2 2} = d^2_{2 M}(\omega_-) \bar{t}_{2 M}= \bar{t}_{2 0} \frac{\sqrt{6}}{4}\sin^2\omega_- - i Im\{\bar{t}_{2 1} \}
\sin \omega_- \nn \\
&- Re \{\bar{t}_{2 1} \} \cos\omega_-\sin\omega_-
\end{eqnarray}
where $\omega$ is the Wigner rotation associated with the boost from the
$\tau$ rest frame to the $e^+ e^-$ C.M. frame which transforms the $\rho$
four-momentum, $k^\prime_\rho= \Lambda k_\rho$. These rotations have the
following expression  
\begin{eqnarray}
\sin \omega = \frac {M_\rho |\vec{p}_\tau| \sin \theta}{M_\tau 
\mid\vec{k}_\rho^\prime \mid}
\;\;\;\; , \;\; \;\;
\cos \omega = \frac {E_\rho E_\rho^\prime M_\tau - Z M_\rho^2}{\mid 
\vec{k}_\rho \mid \mid \vec{k}^\prime_\rho\mid M_\tau}
\label{wignercos}
\end{eqnarray}
with $(Z,\vec{p}_\tau)$ the $\tau$ four-momentum in the $e^+ e^-$ C.M.
frame and $\theta$ the angle between the $\rho$ and the direction 
of the boost in the $\tau$ rest frame.

Now we have the complete angular distributions of the different
chains with final states: 
\begin{itemize}
\item $ (\pi^- \nu_\tau) (\pi^+ \bar{\nu}_\tau)$  
\item $ (\pi^- \pi^0 \nu_\tau) (\pi^+ \bar{\nu}_\tau) + 
(\pi^+ \pi^0 \nu_\tau) (\pi^- \bar{\nu}_\tau)$  
\item $ (\pi^- \pi^0 \nu_\tau) (\rho^+ \bar{\nu}_\tau) + 
(\pi^+ \pi^0 \nu_\tau) (\rho^- \bar{\nu}_\tau) $  
\end{itemize}

Next we are going to find what is the statistical accuracy one can obtain
in the measure of $g_V^b$

\section{Statistical sensitivity}

To estimate the obtainable precision in the measurement of $g_V^b$
in a B-meson facility, we use the formalism of references 
\cite{cramer,franzini,idealerrors}. In these references they call 
``ideal statistical error'' of a parameter $p$ which enters a function 
$f(x,y;p)$ to be determined experimentally, to the error obtained 
from a least-squares-fit to this function with $N$ events. 
To obtain this error, we use that for large number of events, $N$, the 
likelihood function approaches a gaussian. Then, if the function
$f(x,y;p)$ is normalized to one on the physical region, the ideal
statistical error is given by
\beq
\label{ier}
\sigma_p^2 = \frac{1}{N} [\int (\frac{\delta \ln f(x_1,\dots,x_n;p)}
{\delta p})^2 \cdot  f(x_1,...,x_n;p) d x_1 \dots d x_n]^{-1}
\eeq 

The word ``ideal'' stands for the fact that we are not considering the 
efficiency of the detectors, effects of finite experimental resolution and
we assume an ideal distribution of the $N$ events according to $f(x,y;p)$.

In this case our function $f(x,y;p)$ will be the normalized cross section
to the different channels. First we study the reaction $e^-e^+\rightarrow 
\tau^- \tau^+\rightarrow (\pi^- \nu_\tau) (\pi^+ \bar{\nu}_\tau)$. Our 
result is, in this channel,

\beq
\label{tau2}
\sigma_{g_V^b} = \frac{11.1}{\sqrt{N}}
\eeq
where N is the number of $e^- e^+ \rightarrow \tau^- \tau^+ \rightarrow
(\pi^- \nu_\tau)(\pi^+ \bar{\nu}_\tau)$ events. This means that in a B-meson
facility with $10^8\ \Upsilon(1S)$ produced per year one could get a
sensitivity to $g_V^b$ of $6 \cdot 10^{-2}$ only with this channel. 
In Eq. (\ref{final}) is all the information available in the process, 
but not all these observables will be useful for our purposes. In 
particular the P-even, T-even correlations (${\cal C}_{zz}, {\cal C}_{xx},
{\cal C}_{yy}, {\cal C}_{zx}$) get contributions from 
$|T_{\vec{\lambda},\xi}(\gamma)|^2$, 
so they are order 1 and they have no information on $g_V^b$. 
On the other hand, the 
P-odd, T-odd correlations are not as sensitive to this parameter as the 
polarizations, then we can ask whether we can improve the sensitivity by
integrating out some of the final state variables. From Eq. (\ref{final})
we would like to eliminate the P-even correlations maintaining the 
polarizations and P-odd correlations, but unfortunately we can not 
achieve this result integrating out some final state variables.
The only interesting possibility is to consider both $\tau$ decays 
independently, integrating out one of the two pion directions with respect
to the $\tau$. In this case we are considering the $e^- e^+ \rightarrow 
\tau^- \tau^+ \rightarrow ( (\pi^- \nu_\tau) \tau^+ + \tau^-
(\pi^+ \bar{\nu}_\tau) )$ events, where both $\tau$ are hadronically 
reconstructed. By doing this we increase considerably the number of events, 
because now we include also the events in wich the second $\tau$ decays
to $\rho$ and $a_1$. This is approximately
a factor of 5 in the number of events for each tau, but we have two 
independent events for each $\Upsilon$ decay, then the number of events 
increases a factor of 10. The sensitivity one can get with this new decay 
distribution ( we obtain it directly from (\ref{final}) integrating out
$(\theta_+,\phi_+)$ ) is

\beq
\label{tau1}
\sigma_{g_V^b} = \frac{14.6}{\sqrt{N}}
\eeq
and now N is the number of $e^- e^+ \rightarrow \tau^- \tau^+ \rightarrow 
( (\pi^- \nu_\tau)\tau^+ + \tau^- (\pi^+ \bar{\nu}_\tau) )$ events. 
We can see that the difference between Eqs. (\ref{tau2}) and (\ref{tau1})
is roughly a factor $\sqrt{2}$, this is due to the fact that in Eq. 
(\ref{tau2}) we included both the $\tau^+$ and $\tau^-$ polarizations
while in Eq. (\ref{tau1}) we take into account only one of them. On the 
other hand, this means that the P-odd correlations do not improve the
measurement of $g_V^b$ in a significant way. 
 Again with $10^8 \Upsilon$ per year,
one can get a sensitivity of $2.3\cdot 10^{-2}$.
Even more, the simplest polarization analyzer we can use is the 
energy of the pions. The energy of the pions in LAB is related with 
the angle in C.M. of the $\tau$

\beq
E_\pi = \frac{ E_\pi^* E_\tau + q k^*_\pi \cos \theta_-}{M_\tau}
\eeq
where $(E_\pi^*, \vec{k}^*_\pi (\theta_-))$ is the four-momentum of
the pion in  the $\tau$ C.M., and $(E_\tau, \vec{q})$ the four-momentum
of the $\tau$ in LAB. Again from Eq. (\ref{final}), if we integrate 
all the angular variables but $\theta_-$ and make this change of variable
to $E_\pi$ one gets a sensitivity of

\beq
\sigma_{g_V^b} = \frac{20.9}{\sqrt{N}}
\eeq
And, in this way, we do not put any restriction on the second $\tau$ decay, 
then with $10^8 \Upsilon$ per year one can get a sensitivity to $g_V^b$ of $2.7
\cdot 10^{-2}$, simply using the $\pi \nu$ channel and measuring only the 
pion energy.
From this point of view, it is evident that, if we consider only the $\pi \nu$
decay channel, this is the best strategy to measure $g_V^b$, because one can 
use all the $\tau \rightarrow \pi \nu$ events and is experimentally simpler.

We have also studied the channel $\tau \rightarrow \rho \nu$ as a 
polarization analyzer. Then if we apply Eq. (\ref{ier}) to the 
complete distribution $e^-e^+\rightarrow \tau^- \tau^+\rightarrow ((\pi^-\pi^0) \nu_\tau)  
(\rho^+ \bar{\nu}_\tau) + ((\pi^+\pi^0) \nu_\tau)  
(\rho^- \bar{\nu}_\tau) $ given by Eq. (\ref{3steps}) one gets

\beq
\sigma_{g_V^b} = \frac{12.4}{\sqrt{N}}
\eeq
that is slightly worse than our result in the  $e^- e^+ \rightarrow \tau^- \tau^+ 
\rightarrow(\pi^- \nu_\tau)(\pi^+ \bar{\nu}_\tau)$ channel, but now we have more 
events because these channel has a bigger branching ratio. This translates
in a B-meson facility with $10^8 \Upsilon$ produced per year in
a sensitivity to $g_V^b$ of $2.1 \cdot 10^{-2}$.
Other possibility is to use a combined channel as $e^- e^+ \rightarrow 
\tau^- \tau^+ \rightarrow ( ((\pi^-\pi^0) \nu_\tau)(\pi^+ \bar{\nu}_\tau) + 
(\pi^- \nu_\tau)((\pi^+\pi^0) \bar{\nu}_\tau) )$ where the sensitivity to 
the polarization of the $\tau^+\rightarrow \pi^+ \bar{\nu}_\tau$ is better 
than $\tau^+\rightarrow \rho^+ \bar{\nu}_\tau$ if we do not analyze the 
next decay $\rho^+ \rightarrow \pi^+ \pi^0$. The result is,
\beq
\sigma_{g_V^b} = \frac{9.1}{\sqrt{N}}
\eeq
and the number of events is similar to the previous case. The sensitivity 
to $g_V^b$ that one can reach in this channel, with $10^8 \Upsilon$, is 
$2.3 \cdot 10^{-2}$. As in the $\pi$
channel we can increase the statistics by integrating one of the decays,
although we lose some sensitivity. We integrate the direction of one of 
the $\rho$ or equivalently the $\pi$ direction in Eq. (\ref{3steps}) 
and only require this $\tau$ to decay hadronically. Then we keep simply
the decay of a $\tau$ to $\rho \nu$ and then the decay of $\rho$ to 
$\pi \pi$. Then we get a sensitivity of,
 
\beq
\sigma_{g_V^b} = \frac{14.6}{\sqrt{N}}
\eeq
but now the number of events has increased a factor of 10, because N
is the number of $e^- e^+ \rightarrow \tau^- \tau^+ 
\rightarrow ( ((\pi^- \pi^0) \nu_\tau)\tau^+ + \tau^- 
((\pi^+ \pi^0) \bar{\nu}_\tau) )$ events. Again with $10^8 \Upsilon$ per year,
one gets a sensitivity of $1.6\cdot 10^{-2}$.

After this analysis we can combine a series of independent measurements
into a final value for $g_V^b$ with an error given by

\beq
\label{sumaerr}
\sigma=(\sum_i \frac{1}{\sigma_i^2})^{-1/2}
\eeq 
  
then we can combine the error obtained with the channels $e^-e^+
\rightarrow (\pi^- \nu_\tau) (\pi^+ \bar{\nu}_\tau)$, $e^-e^+
\rightarrow (((\pi^-\pi^0) \nu_\tau) (\rho^+ \bar{\nu}_\tau) +((\rho^- \nu_\tau)
((\pi^+ \pi^0) \bar{\nu}_\tau)) $ and  $e^- e^+ \rightarrow 
( ((\pi^- \pi^0) \nu_\tau)(\pi^+ \bar{\nu}_\tau) + (\pi^- \nu_\tau)
((\pi^+ \pi^0) \bar{\nu}_\tau) )$ and with $10^8 \Upsilon$ one gets a 
sensitivity to $g_V^b$ of $1.5\cdot 10^{-2}$.

On the other hand, we can also combine the errors obtained in the  $e^- e^+ 
\rightarrow ( ((\pi^- \pi^0) \nu_\tau)\tau^+ + \tau^-
((\pi^+ \pi^0) \bar{\nu}_\tau) )$ and  $e^- e^+ 
\rightarrow ( (\pi^- \nu_\tau)\tau^+ + \tau^- (\pi^+ \bar{\nu}_\tau) )$ 
channels, and again with $10^8 \Upsilon$ one gets a sensitivity of 
$1.3\cdot 10^{-2}$.

Notice this result has been obtained with a sample of $10^8 \Upsilon$, 
for a different number of $\Upsilon$ produced, the sensitivity to
$g_V^b$ would simply re-scale by a factor $\sqrt{10^8/N_\Upsilon}$.

\section{Conclusions} 

In this work we have studied the possibilities of a high luminosity 
B-meson facility to measure with high precision the $Z- b \bar{b}$
vector coupling. At the energies of $\Upsilon (1S)$ we have used the 
$\tau^- \tau^+ $ channel to determine this coupling through the $\tau$
polarizations. A complete analysis of the hadronic decay modes of
the $\tau$ lepton has been done, with special attention to the
$\tau^- \rightarrow \pi^- \nu_\tau$ and $\tau^- \rightarrow \rho^- \nu_\tau$ 
as polarization analyzers. We have built the complete correlated
cross section with the decays of both $\tau$s and from here we have
calculated the ideal statistical errors obtainable in the measure of
$g_V^b$. We have found that in a one year run in a B-meson facility 
with $10^8 \Upsilon$ per year, one can get a sensitivity of $1.3\cdot 10^{-2}$,
comparable with the present precision in this coupling from the LEP/SLC 
measurements of $R_b$ and $A_b$.

\section*{Acknowledgments}

O.V. acknowledges the Generalitat Valenciana for a research fellowship. 
This work has been supported by Grant AEN96-1718 of the Spanish CICYT.

\newpage
\section*{APPENDIX A}
\setcounter{equation}{0}
\def\theequation{A.\arabic{equation}}

In a decay, $A(\sigma) \rightarrow B(\lambda_b) + C(\lambda_c)$, we can 
obtain the density matrix, 
$\rho^{out}$, describing the complete final state of particles B and C in 
terms of helicity amplitudes and the initial density matrix in the 
following way,
\beq
\label{2density}
\rho^{out}_{\vec{\lambda},{\vec{\lambda^\prime}}} = \sum_{\sigma \sigma^\prime}
f_{\vec{\lambda},\sigma}\ (\Omega_1)\ \rho^{in}_{\sigma \sigma^\prime} 
(\Omega)\ f_{\vec{\lambda^\prime},\sigma^\prime}^*\ (\Omega_1)
\eeq
where $\vec{\lambda}= (\lambda_b, \lambda_c)$.

It is very convenient to express our initial density matrix in a 
basis of irreducible tensor operators, $T_{L,M}$, with coefficients $t_{L,M}$,
\cite{phi,martin},

\begin{equation}
\label{irr.tensor}
\frac{\rho^{in}}{Tr(\rho^{in})} = \frac{1}{2j+1} \sum_{L,M}^{2j} (2L+1)
t_{L,M}^{(a)\ *} T_{L,M}^{(a)}
\end{equation}
these coefficients, $t_{L,M}^{(a)}$, are the so-called multipole parameters,
\begin{equation}
\label{mult.param}
Tr(\rho^{in}) t_{L,M}^{(a)}(\theta) = Tr (\rho^{in} T_{L,M}^{(a)}) =
\sum_{\sigma, \sigma^\prime} (\rho^{out})_{\sigma, \sigma^\prime}
C(1 L 1 | \sigma M \sigma^\prime)
\end{equation}
where these $C(j L j|m^\prime M m)$ are Clebsch-Gordan coefficients.
 
For $L=1$, we can relate, \cite{martin}, the usual polarizations
and the multipole parameters,
\begin{equation}
P_{x^\prime} = - (t_{1,1} - t_{1,-1}) = -2\ Re[ t_{1,1}]
\end{equation}
\begin{equation}
P_{y^\prime} = i (t_{1,1} + t_{1,-1}) = -2 \ Im[t_{1,1}]
\end{equation}
\begin{equation}
P_{z^\prime} = \sqrt{2}\ t_{1,0}
\end{equation}

Using Eq. (\ref{irr.tensor}) to express the initial density matrix in the 
basis of irreducible tensors and replacing the helicity amplitudes in the
C.M. frame of the decaying particle with Eq. (\ref{ampldec}), we get the 
final density matrix in terms of the multipole parameters of the $A$ particle
and the reduced helicity amplitudes,
\bea
\label{red.dens}
\rho^{out}_{\vec{\lambda},\vec{\lambda^\prime}} &= &Tr(\rho^{in})\,
\frac{\sqrt{2j+1}}{4 \pi} \sum_{L,M}\, \sqrt{2L+1}\ (-)^{j-\lambda^\prime}\ 
T_{\vec{\lambda}}\,T_{\vec{\lambda^\prime}}^* \\
&& \ t_{L,M}^{(a)\ *}\, 
C(j, j, L|\lambda^\prime, -\lambda, \lambda^\prime-\lambda)\, 
{\cal D}^{(L) *}_{M,\lambda-\lambda^\prime} (\Omega)
\nn
\eea
This density matrix contains all the information available in the process.
For instance, if we want the angular distribution we just have to take 
the trace on $\vec{\lambda}$, and in the same way the density matrix 
for one of the final particles is obtained taking the trace on the 
helicities of the other particle.

To define a complete set of observables we generalize Eq. (\ref{irr.tensor})
\beq
\label{gen.tensor}
\frac{\rho^{out}}{Tr(\rho^{out})} = \frac{1}{(2j_1+1)(2j_2+1)}
 \sum_{L,M,L^\prime,M^\prime}^{2j} (2L+1) (2L^\prime+1) 
{\cal C}_{L,M,L^\prime,M^\prime}^* 
T_{L,M}^{(b)} T_{L^\prime,M^\prime}^{(c)}
\eeq
and then these generalized multipole parameters, 
${\cal C}_{L,M,L^\prime,M^\prime}$, include all the information on the 
density matrices of particles B and C, 
$t_{L,M}^{(b)}={\cal C}_{L,M,0,0}$, $t_{L,M}^{(c)}={\cal C}_{0,0,L,M}$ 
and additionally the correlations between them.

It is very important to notice here that Eq. (\ref{red.dens}) is only valid
in the C.M. frame of the decaying particle. However, in general in the LAB 
frame the decaying particle will move with a 
momentum different from zero. So, we will be interested in the transformation
properties of these density matrices under Lorentz boosts.

The transformation of an helicity amplitude under a boost from the C.M. 
frame to the LAB frame is just a rotation, the so-called Wigner rotation
\cite{wigner,wick,boost}, that if we choose $\phi=0$, is \cite{martin},
\bea
\label{wigamplitudes}
f^{(CM)}_{\lambda_b \lambda_c,\sigma}(\theta,\phi=0) & = & 
\sum_{\lambda_b^{\prime} 
\lambda_c^{\prime}}(-)^{\lambda_c^{\prime} - \lambda_c}\; 
d^{j_b}_{\lambda_b^{\prime} \lambda_b} (\omega_b)\; 
d^{j_c}_{\lambda_c^{\prime} \lambda_c} (\omega_c)\;  
f_{\lambda_b^\prime \lambda_c^\prime,\sigma}^{(LAB)}(\theta,\phi=0) 
\nn 
\\
\eea
where this rotation, of angle $\omega_i$, is given by \cite{martin},
\bea
\sin \omega_i= \frac {M_i \sinh \kappa \sin \theta}{
\mid\vec{p}_i^\prime \mid}
\;\;\;\; , \;\; \;\;
\cos \omega_i = \frac {E_i^\prime E_i - \cosh \kappa M_i^2}{\mid 
\vec{p}_i\mid  \mid \vec{p}_i^\prime\mid}
\label{wigners}
\eea
where $\kappa$ is the parameter of the boost from the C.M. frame of the 
decaying particle to the LAB, related to the velocity by
$v = \tanh\kappa$. $M_i$ the mass of the particle and 
$(E_i,\vec{p_i}(\theta))$ its four-momentum in C.M. of the decaying particle,
transformed under the boost as $p_i^\prime = \Lambda p_i$, in a frame where 
the boost is along the z-axis.
From Eq. (\ref{wigners}) we can see that under a boost collinear to the
particle three-momentum, $\theta=0$, our states do not suffer any
rotation.

The helicity amplitudes in Eq. (\ref{wigamplitudes}) are functions of two
variables. For instance, we could choose the invariant variables $(s,t,u)$, 
keeping the same expression in any frame. Nevertheless, as
we have seen in Eq. (\ref{ampldec}), these helicity amplitudes have a 
specially simple form in terms of C.M. variables. Then, we have used
this freedom to express, both the C.M. and the LAB helicity amplitudes
in Eq. (\ref{wigamplitudes}) in terms of C.M. variables.     

As can see in Eq. (\ref{2density}), density matrices are a product of 
two helicity amplitudes, this means that using Eq. (\ref{wigamplitudes}) 
we can get the transformation of the density matrix of particle $B$
\beq
\label{rotja}
\rho^{LAB} = d^{j_b} (\omega_b) \cdot \rho^{CM} \cdot d^{j_b\ T} (\omega_b)
\eeq

It is very interesting to see how the multipole parameters are affected
by these rotations. We use Eq. (\ref{mult.param}) to express $\rho^{CM}$
in terms of these multipole parameters

\bea
\label{rot.param}
\rho^{LAB}_{\sigma \sigma^\prime}&= \frac{1}{2j+1}\sum_{L M} (2L+1) 
\sum_{\lambda\lambda^\prime} d^{(j)}_{\sigma \lambda} (\omega_a) 
(j L j | \lambda^\prime M \lambda)
d^{(j)\ T}_{\sigma^\prime\lambda^\prime} (\omega_a) t^{(b)\ *}_{L M} \nn\\ 
&= \frac{1}{2j+1}\sum_{L M} (2L+1)  (j L j | \sigma^\prime \sigma-
\sigma^\prime) d^{(L)\ T}_{\sigma-\sigma^\prime M} (\omega) t^{(b)\ *}_{L M}
\eea
and comparing again with Eq. (\ref{mult.param}) that in LAB we get a new 
set of multipole parameters that are obtained simply applying the rotation 
to the C.M. ones. 
\beq
\label{rotmulti}
t_{L M}^{(b)\ *} = \sum_{M^\prime} d^L_{M M^\prime}(\omega_b) 
t_{L M^\prime}^{(b)\ *} 
\eeq
This is all we need to obtain the multipole parameters and density 
matrices in the LAB frame.

\section*{APPENDIX B}
\setcounter{equation}{0}
\def\theequation{B.\arabic{equation}}

In this appendix, we present the general method to calculate reduced 
helicity amplitudes, and we apply it to some examples in the
$e^+e^- \rightarrow \tau^+ \tau^-$ processes. 

Reduced helicity amplitudes are easily calculable by means of Eq. 
(\ref{ampli 2}) from the helicity amplitudes. So, our first step will 
be to obtain these helicity amplitudes from the Feynman amplitudes 
we can calculate from the diagrams with Feynman rules. Taking into account the 
normalization we have defined for our reduced helicity amplitudes in Eqs. 
(\ref{ampli 2}) and (\ref{ampldec}) the difference between them and the
Feynman amplitudes will just be a $q^2$-dependent phase space factor, 
irrelevant in all our observables. So, we can simply define:
\beq
\label{irrelevant}
M_{\sigma,\lambda_+,\lambda_-} (\theta) = K f_{\sigma,\vec{\lambda}}(\theta)
\eeq
with $M_{\sigma,\lambda_+,\lambda_-} (\theta)$ the Feynman amplitudes.

Now we will explicitly apply this procedure to the calculation of
the reduced helicity amplitudes $T_{(\lambda,\lambda^\prime),\xi}(\gamma)$ and 
$T_{(\lambda,\lambda^\prime),\xi}(\gamma Z_A)$ corresponding to diagrams
$(1) + (2)$ and $(3) + (5)$ in Figure 1.

The Kinematics in the C.M. frame of the $e^+ e^-$ system is defined by

\begin{eqnarray}
\label{kinematics}
l^\mu_- = (E,0,0,|\vec{l}|) & q^\mu = (l_-+l_+)^\mu \\
l^\mu_+ = (E,0,0,-|\vec{l}|) & k^\mu = (k^0,|\vec{k}|\sin\theta,0,
|\vec{k}|\cos\theta) \nonumber 
\end{eqnarray}
where $l_\pm^\mu$ is the four-momentum of the $e^\pm$, whose helicities
are $\xi_\pm=\pm 1/2$, $k^\mu$ the
four-momentum of the $\tau^-$ and the helicities of the $\tau^\pm$
will be denoted as $\lambda_\pm= \pm 1/2$.

The Feynman amplitudes corresponding to these diagrams are

\begin{eqnarray}
\label{gammatau}
M^{\gamma}_{\lambda_-,\lambda_+,\xi_+,\xi_-} (\theta) &=& i \frac{e^2}{s}
(1 + \frac{e^ 2}{s} Q_b^2 |F_\Upsilon(q^2)|^2 P_\Upsilon(q^2))  
V_{\nu}^\tau(\lambda_-,\lambda_+,\theta) g_{\mu \nu} V_\mu^{e *}(\xi_-,\xi_+) 
\nn \\
\end{eqnarray}
\begin{eqnarray}
\label{gamma-Z}
M^{\gamma Z_A}_{\lambda_-,\lambda_+,\xi_-,\xi_+} (\theta) &=& i \frac{8 G_F}{\sqrt{2}}
g_A (\frac{e^2}{s} Q_b g_V^b |F_\Upsilon(q^2)|  P_\Upsilon(q^2) - g_V)
\nn \\
&&A_\nu^\tau(\lambda_-,\lambda_+,\theta) g_{\mu \nu} V_\mu^{e *}(\xi_-,\xi_+)
\end{eqnarray}
where we have followed the notation of sections 2, and we 
have introduced the matrix elements of the leptonic currents, 

\begin{eqnarray}
\label{lept.curr}
&V^\mu_{l}(\lambda_-,\lambda_+) = \bar{u}(p_-,\lambda_-) \gamma^\mu 
v(p_+,\lambda_+) \nonumber \\
&A^\mu_l(\lambda_-,\lambda_+) = \bar{u}(p_-,\lambda_-) \gamma^\mu \gamma_5 
v(p_+,\lambda_+)
\end{eqnarray}

Now we need to obtain an explicit expression for these matrix elements. 
To do this, we follow the method of reference \cite{fearing}, which permits
the calculation of these amplitudes using standard trace techniques. 
Then, the complete results for the vector and axial currents, in the 
CM frame and with the momenta along the $z$-axis are,
\begin{eqnarray}
\label{currents}
&V^\mu_{l}(\lambda_-,\lambda_+=\lambda_-) = (0, 0, 0, -2 M_l)\nn \\  
&V^\mu_{l}(\lambda_-,\lambda_+=-\lambda_-) = (0, 4 E_p \lambda_-,- 2 E_p i, 0)
\nn \\  
&A^\mu_{l}(\lambda_-,\lambda_+=\lambda_-) = (-4 M_l \lambda_-, 0, 0, 0)\nn \\  
&A^\mu_{l}(\lambda_-,\lambda_+=-\lambda_-) = (0, 2 p, -4 i p \lambda_-, 0)  
\end{eqnarray}
With these results we have all the leptonic currents we need, because 
they are perfectly behaved Lorentz vectors or axial-vectors. Then, we 
just have to rotate them, if the momentum is in a different direction. 

With all these elements, we simply use Eq. (\ref{irrelevant}) and Eq. 
(\ref{ampli 2})
with our Feynman amplitudes, Eqs. (\ref{gammatau}) and (\ref{gamma-Z}),
to obtain the reduced helicity amplitudes. For instance, the expression
for $M^{\gamma}_{\vec{\lambda}=(1/2,1/2),\vec{\xi}=(1/2,-1/2)}(\theta)$ is,
\begin{eqnarray}
\label{gammatau.sus}
M^{\gamma}_{(1/2,1/2),(1/2,-1/2)} (\theta) = i \frac{e^2}{s}
(1 + \frac{e^ 2}{s} Q_b^2 |F_\Upsilon(q^2)|^2 P_\Upsilon(q^2))\nn \\  
(0,-2M_\tau \sin \theta,0,-2M_\tau \cos \theta)\cdot
(0,-2\frac{\sqrt{s}}{2},-2i\frac{\sqrt{s}}{2},0)^T = \nn \\
i \frac{e^2}{s}(1 + \frac{e^ 2}{s} Q_b^2 |F_\Upsilon(q^2)|^2 P_\Upsilon(q^2))
4 M_\tau \frac{\sqrt{s}}{2} \ (-\sqrt{2})\ (\frac{-\sin \theta}{\sqrt{2}})    
\end{eqnarray}
where we have applied a rotation to the leptonic current of the $\tau$
with respect to Eq. (\ref{currents}) and we have taken the complex
conjugate of Eq. (\ref{currents}) to obtain the electron current. The extra 
minus sign is due to the metric $g_{\mu \nu}$. In Eq. (\ref{gammatau.sus})
we just have to remove the rotation matrix element $d_{1 0}(\theta)$, which
is exactly the last term in this equation. This procedure has to be
repeated with all the amplitudes and then, finally we get the 
following results

\begin{eqnarray}
\label{redgammatau01}
K T_{(+,+),1}(\gamma)= -i 4 \sqrt{2}\ \frac{e^2}{s}\ (1 +\frac{e^2}{s} 
Q_b^2 |F_\Upsilon|^2 P_\Upsilon)\ M_\tau\ \frac{\sqrt{s}}{2}
\end{eqnarray}

\beq
\label{redgammatau11}
K T_{(+,-),1}(\gamma)= -i 8\ \frac{e^2}{s}\ (1 + \frac{e^2}{s} Q_b^2 
|F_\Upsilon|^2 P_\Upsilon)\ p^0\ \frac{\sqrt{s}}{2}
\eeq

\beq
\label{redgammaZ01}
K T_{(+,+),1}(\gamma Z_A)= 0 
\eeq

\beq
\label{redgammaZ11}
K T_{(+,-),1}(\gamma Z_A)= -i 8\ \frac{8 G_F}{\sqrt{2}}\ g_A\ 
(\frac{e^2}{s} Q_b g_V^b |F_\Upsilon|^2 P_\Upsilon-g_V)\ |\vec{p}|\ 
\frac{\sqrt{s}}{2} 
\eeq
In the same way, we can obtain all the reduced helicity amplitudes in
Eq. (\ref{amplred}).




\newpage
\hspace{-0.6cm}{\Large \bf Figure Captions} 
\vspace{2cm}
\\
\\
\\
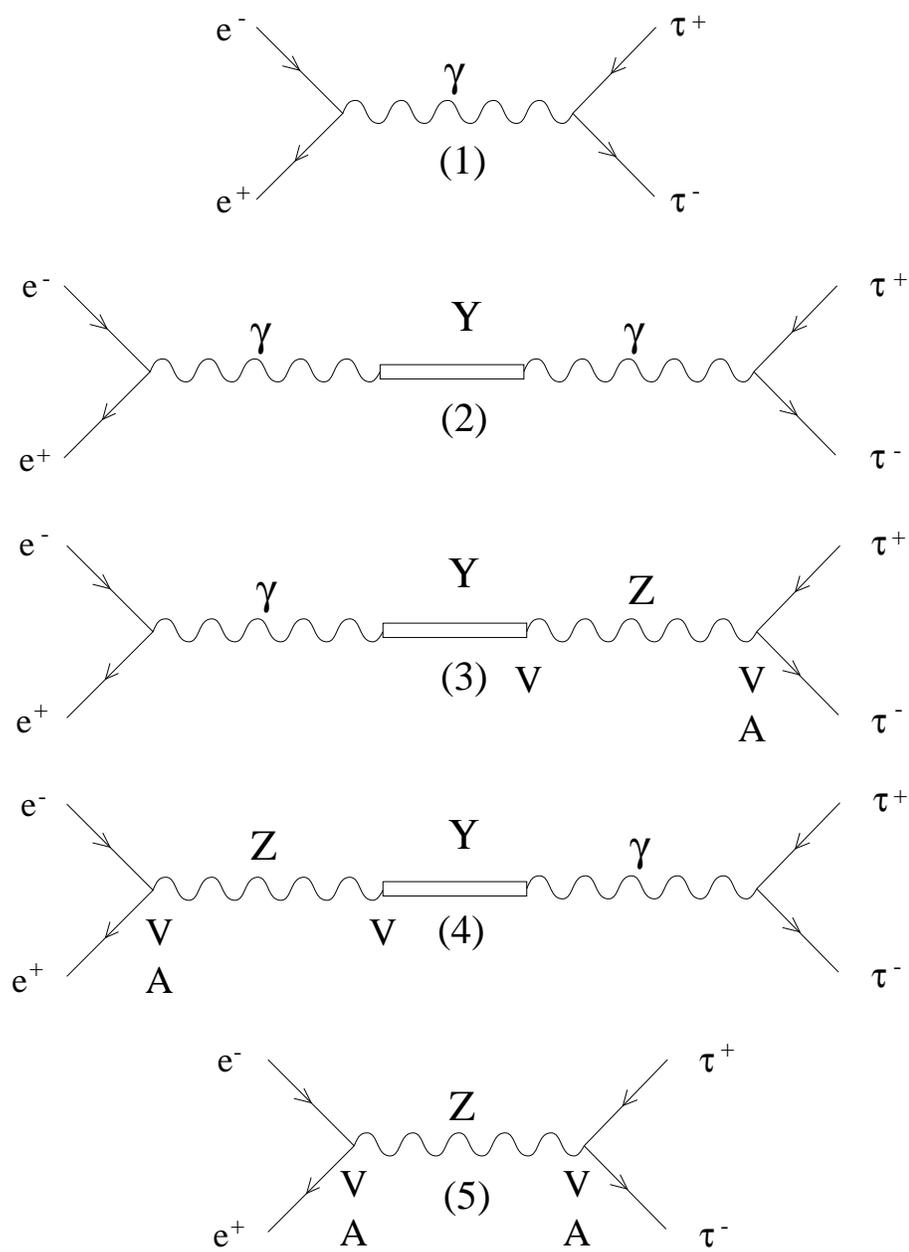
{\bf Figure 1}: Resonant and non-resonant diagrams for the process $e^+e^-
\rightarrow \tau^+ \tau^-$  at the $\Upsilon$ region.
\\
\\
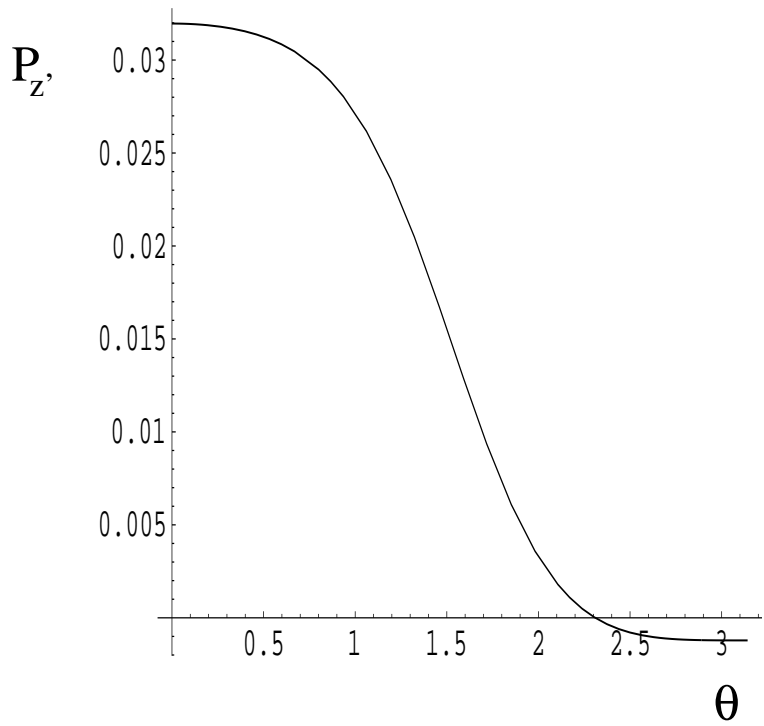
{\bf Figure 2}: Longitudinal $\tau^-$ polarization, $P_{z'}(\theta)$
\\
\\
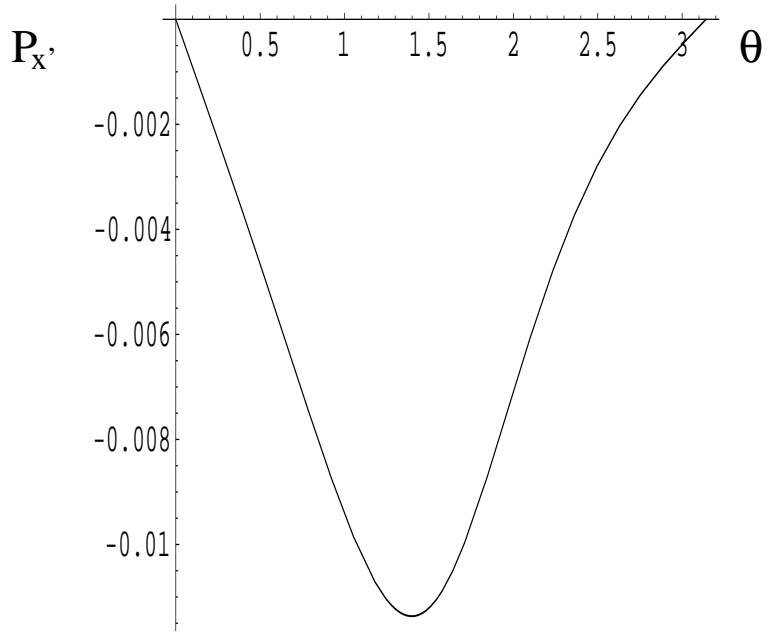
{\bf Figure 3}: Transverse $\tau^-$ polarization, $P_{x'}(\theta)$

\newpage
\hspace*{-0.6cm}{\Large \bf Table Captions}\\
\\
\\
\\
{\bf Table 1}: Dominant $\Upsilon(1S)$ decay channels. 
\\
\\
{\bf Table 2}: Dominant $\tau$ decay channels. 
\newpage
\begin{center}
{\bf Table 1}
\end{center}
\begin{table}[t,h,b]
\begin{center}
\begin{tabular}{|l|l|}
\hline
\multicolumn{2}{|c|}{$\Upsilon (1S)\; \; \;\;  I^G(J^{PC})=0^-(1^{--})$}\\
\hline
\multicolumn{2}{|c|}{Mass $\;M_\Upsilon\; = 9460.37 \pm 0.21\; MeV $}\\
\multicolumn{2}{|c|}{Width $\;\Gamma\;= 52.5 \pm 1.8\;KeV $}\\ \hline
Decay modes  &  Fraction $\Gamma_i /\Gamma $  \\ \hline
 $ \tau^+ \tau^- $ & $(2.97 \pm 0.35) \% $  \\
 $ \mu^+ \mu^- $ & $(2.52 \pm 0.17) \% $  \\
 $ e^+ e^- $ & $(2.48 \pm 0.07) \% $  \\
 $ J/\psi(1S)\; anything  $ & $ (1.1 \pm 0.4) \times 10^{-3} $\\
 $\gamma\; 2h^+ 2h^-$ & $ (7.0 \pm 1.5) \times 10^{-4}  $\\
 $\gamma\; 3h^+ 3h^-$ & $ (5.4 \pm 2.0) \times 10^{-4}  $\\
 $\gamma\; 4h^+ 4h^-$ & $ (7.4 \pm 3.5) \times 10^{-4}  $\\
 \hline
\end{tabular}
\end{center}
\end{table}

\begin{center}
{\bf Table 2}
\end{center}
\begin{table}[t,h,b]
\begin{center}
\begin{tabular}{|l|l|}
\hline
\multicolumn{2}{|c|}{$\tau\; \; \;\;  J =\frac{1}{2}$}\\
\hline
\multicolumn{2}{|c|}{Mass $\;M_\tau\; = 1777.00^{+0.30}_{-0.27} \; MeV $}\\
\multicolumn{2}{|c|}{Mean life $\;\tau\;= (291.0 \pm 1.5) \times 10^{-15}\;s$}
\\ \hline
Decay modes  &  Fraction $\Gamma_i /\Gamma $  \\ \hline
 $ \mu^- \bar{\nu}_\mu \nu_\tau $ & $(17.35 \pm 0.10) \% $  \\
 $ e^- \bar{\nu}_e \nu_\tau $ & $(17.83 \pm 0.08) \% $  \\
 $ \pi^- \nu_\tau $ & $(11.31 \pm 0.15) \% $  \\
 $ \pi^- \pi^0 \nu_\tau  $ & $ (25.24 \pm 0.16) \% $\\
 $ h^- 2\pi^0 \nu_\tau$ & $ (9.50 \pm 0.14) \%  $\\
 $ h^- h^- h^+ \nu_\tau $ & $ (9.80 \pm 0.10) \%  $\\
 \hline
\end{tabular}
\end{center}
\end{table}

\end{document}